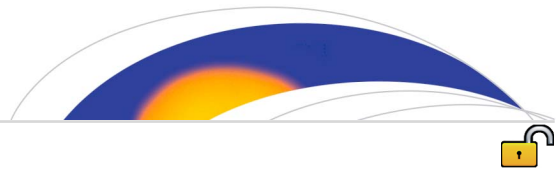

## Space Weather

# Solar Sources of Interplanetary Magnetic Clouds Leading to Helicity Prediction

**Roger K. Ulrich[1]** , **Pete Riley[2]** , **and T. Tran[1]**

[1]Department of Physics and Astronomy, University of California, Los Angeles, CA, USA, [2]Predictive Science Incorporated, San Diego, CA, USA

**Abstract** This study identifies the solar origins of magnetic clouds that are observed at 1 AU and predicts the helical handedness of these clouds from the solar surface magnetic fields. We started with the magnetic clouds listed by the Magnetic Field Investigation (MFI) team supporting NASA's Wind spacecraft in what is known as the MFI table and worked backward in time to identify solar events that produced these clouds. Our methods utilize magnetograms from the Helioseismic and Magnetic Imager instrument on the Solar Dynamics Observatory spacecraft so that we could only analyze MFI entries after the beginning of 2011. This start date and the end date of the MFI table gave us 37 cases to study. Of these we were able to associate only eight surface events with clouds detected by Wind at 1 AU. We developed a simple algorithm for predicting the cloud helicity that gave the correct handedness in all eight cases. The algorithm is based on the conceptual model that an ejected flux tube has two magnetic origination points at the positions of the strongest radial magnetic field regions of opposite polarity near the places where the ejected arches end at the solar surface. We were unable to find events for the remaining 29 cases: lack of a halo or partial halo coronal mass ejection in an appropriate time window, lack of magnetic and/or filament activity in the proper part of the solar disk, or the event was too far from disk center. The occurrence of a flare was not a requirement for making the identification but in fact flares, often weak, did occur for seven of the eight cases.

**Plain Language Summary** We started with clouds of ionized gas threaded by a helical magnetic field and embedded in the solar wind near Earth that were identified and modeled by the Wind team. We then worked backward in time to find solar events as sources for these clouds. We were able to find these sources for only 8 of 37 candidate clouds. Finding the location of the cloud source required us to define what we call *magnetic origination points* where we evaluated the field configuration. We used two methods to predict the handedness of these clouds. One method predicted all eight correctly, and the other method predicted seven correctly.

## 1. Introduction

Solar outbursts of varying sizes and properties can produce effects on Earth and within the Heliosphere that are generally known as space weather. The size of the effect can vary from barely detectable to large enough to have substantial societal impact. In 2008 the National Academy of Science provided a summary of severe space weather events (Committee on the Societal and Economic Impacts of Severe Space Weather Events: A Workshop, National Research Council, 2008). As explained in that report, the prediction of the strength of such events can mitigate their disruption and cost. For many years the coincidence between solar flares and geomagnetic storms suggested a causal relationship (Carrington, 1859; Hale, 1931). Geomagnetic storms were known to be associated with a variety of solar wind configurations such as magnetic clouds and compound streams (now referred to as *corotating interaction regions*; Burlaga et al., 1987), but the origin was still considered to be solar flares (Baker et al., 1984). In fact, the identification of solar sources of these solar wind configurations was substantially clarified when Gosling (1993) provided convincing arguments that solar flares are sometimes associated with the occurrence of geomagnetic storms but are neither necessary nor sufficient conditions for these storms. The magnetic clouds are now accepted as coming from what are called *coronal mass ejections* or CMEs.

In that seminal paper Gosling (1993) displaced the central role of solar flares in the generation of interplanetary CMEs (ICMEs) and proposed a now generally accepted picture of the development of CMEs in the solar corona and their subsequent geomagnetic effects. Later, Bothmer and Schwenn (1998) studied the structure of magnetic clouds observed by Helios 1 and 2 spacecraft, relating them to the properties of associated large quiescent filament disappearances and finding a general agreement with the helicity patterns of both. A number of studies have attempted to uncover the relationship between ICMEs and their solar origins (see reviews by, e.g., Burlaga, 2002; Linker et al., 2003). Some other studies have focused on white-light images from space-born coronagraphs to establish the connections between solar and ICMEs. For example, Riley et al. (2008) related the three-part structures of CMEs consisting of (1) a bright front, (2) a dark cavity, and (3) a bright, compact core as observed in the solar corona to their in situ counterparts. However, the connection is often not clear as was found by Nitta and Mulligan (2017) who described a set of CMEs that were not accompanied by any obvious low coronal signatures, the so-called stealth events. Many case studies have also been presented. Harrison et al. (2012), for example, investigated the 2010/08/01 (we give dates as yyyy/mm/dd) eruptions describing the complex eruption, propagation, and evolution of four major CMEs. Other studies have attempted to connect the magnetic properties of features in the low corona and/or photosphere with magnetic clouds such as that by Marubashi et al. (2017) who compared the orientation of interplanetary flux ropes with the orientation of the magnetic polarity inversion line (PIL) for a selection of events, inferring that their flux-rope structures were likely created in the corona.

As stated by Gonzalez and Tsurutani (1987), the dominant parameter responsible for the development of the main phase of geomagnetic storms is the southward component of the interplanetary magnetic field ($-B_z$ in solar magnetospheric coordinates). The analysis by Klein and Burlaga (1982) showed that the magnetic configuration in many magnetic clouds arriving at Earth is helical in structure with the transverse magnetic field vector rotating through an arc. The sense of the helicity governs the initial value of the southward field and its subsequent evolution. Bothmer and Schwenn (1998) discussed the helical magnetic structure using a cylindrical, force-free flux tube configuration. Using a magnetic field model based on this structure, which was given by Lepping et al. (1990), the Magnetic Field Investigation (MFI) team supporting NASA's Wind spacecraft (Lepping et al., 1995) prepared an extensive listing of identified magnetic clouds and their properties covering the years 1995 to 2012 (https://wind.nasa.gov/mfi/mag_cloud_S1.html), which we refer to as the *MFI table*. The listed clouds have been modeled in a uniform manner, and all have derived helicity, which we can use to test the validity of helicity predictions based on solar surface magnetic field observations. For this purpose we need to identify the solar origins of clouds on the MFI list.

Solar active region structures are generally believed to result from flux tubes that have emerged from the solar interior (Schrijver, 2009). The pattern of magnetic field in these active regions often is simple when the region is young and appears as a bipolar spot group. As the region matures or ages, it becomes more complex with a multipolar magnetic pattern or with small sections of polarity opposite to that of the nearby solar surface. NASA's Solar Dynamics Observatory (SDO) carries two investigations that we use extensively to understand the structure of active regions and the relationship between the magnetic pattern and hot regions in the low corona. These investigations are the Helioseismic and Magnetic Imager (HMI; Scherrer et al., 2012) and the Atmospheric Imaging Assembly (AIA; Lemen et al., 2012). HMI provides the solar surface magnetic field measurements, while AIA provides images of the chromosphere, transition region, and the low corona using four telescopes with seven EUV channels and three longer wavelength channels. In our use of these images we designate them as AIA-XXX where XXX is the filter wavelength in Å. We also use the CDAW (Coordinated Data Analysis Workshop) catalogue maintained by the LASCO (Large Angle and Spectrometric Coronagraph) team to identify the CMEs as they leave the vicinity of the sun. We primarily use images from the LASCO/C2 telescope, which is on board the Solar and Heliospheric Observatory spacecraft. We consider those CMEs that leave the sun in the form of a ring (halo) or nearly complete ring (partial halo) to be candidates for reaching 1 AU near Earth.

Our objective of making a helicity prediction requires identifying the magnetic cloud with the CME and with surface activity, it also requires making an identification of the solar surface magnetic fields from which the cloud originates. The work by Li et al. (2014) has carried out the first of these objectives but did not identify the magnetic originating configuration nor did it address the helicity question. Their study spanned a longer period of time than is included here, and their identifications and ours agree for a few cases but not for others primarily due to our specific requirements for making an identification.

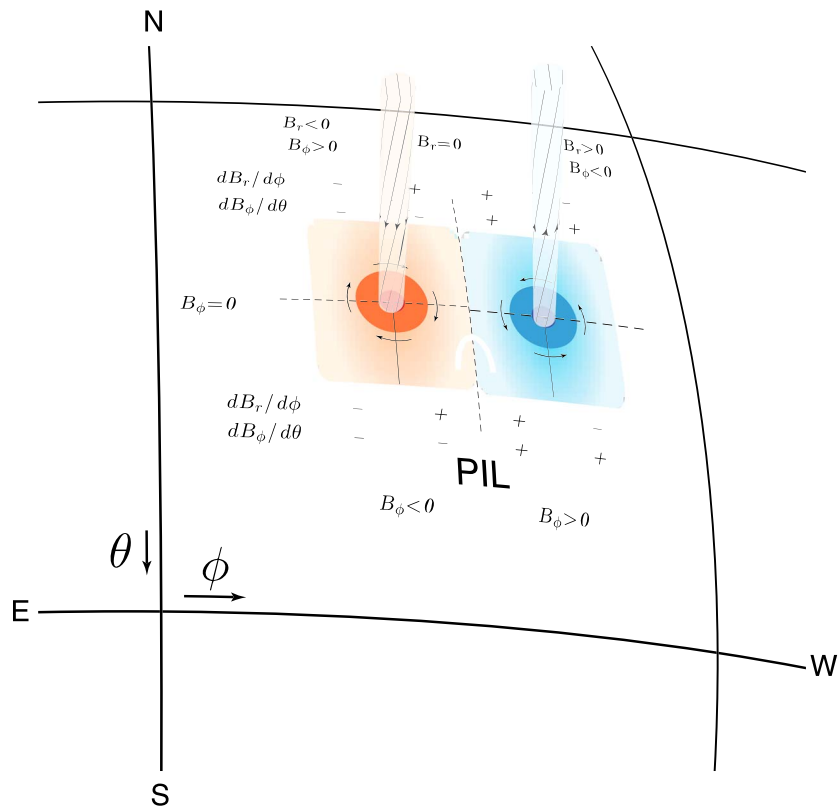

**Figure 1.** This sketch shows the geometry of a flux rope as it crosses the photosphere. The helical direction is shown on the surface of the cylinder for the case of a right-handed helix. The sketch of the magnetic flux rope is schematic and not derived from any model. The sense of the field gradients are shown in each quadrant around the strongest field point. The angles indicated are for spherical coordinates with North at the top and $\theta$ increasing from 0° at the North Pole to 180° at the South Pole. The signs of the derivatives in each of the quadrants indicate the helicity for this RH case. For a left-handed helix, the signs of the transverse quantities would be reversed or the signs of the vertical magnetic fields would be reversed. A convenient way to check both the signs of the derivatives and the signs of the magnetic fields is to examine the product of the vertical field component and the derivative. In section 2.2.3 we define this as $IH$, which is proportional to the current helicity density. A right-handed helix has $IH > 0$, and a left-handed helix has $IH < 0$. The boundary between the upward and downward vertical magnetic field is called the *polarity inversion line* (PIL) and is indicated on this sketch. Flares are often associated with structures near the PIL and the arches like the one indicated here are part of the flare process.

The magnetic field threading flux tubes are often in the form of a helical structure forming a flux rope (see, e.g., Cheng et al., 2017; Hu et al., 2014), which leaves the solar vicinity in the form of a CME and eventually reaches Earth. Figure 1 illustrates the pattern of magnetic field near the points where the flux tube emerges and gives conditions on the angular derivatives for the case of right-handed helicity that can be applied to the left-handed case by either reversing the sign of $B_r$ or rotating the transverse field by 180°. Due to flux conservation, the flux rope should be a loop connecting two points of opposite magnetic field polarity. There are many loops present in active regions so that identification of the ejected loop leading to the CME is an essential first step. The maps of the magnetic fields and their gradients are sometimes complex so that the predicted helicity depends sensitively on the emergent point locations. Consequently, the determination of the positions of the paired emergent points is essential to the prediction of the helicity of the ejected cloud carrying the flux rope. We call these paired points the *magnetic origination points* (MOPs), and the condition that they have opposite polarity often dictates which locations can be identified.

PILs are boundaries between areas of opposite sign of the radial magnetic field and often have low-lying arcades of arches going over the PIL from one polarity to the other. These often are heated during the flare but are not ejected. Associated with flares are features known as *foot points*, which are parts of loops near the site of a flare that undergo dimming due to the loss of loop plasma. These foot points or dimming regions are commonly considered to be the points of origin of CMEs (Cheng & Qiu, 2016; Cheng & Ding, 2016;

Chertok et al., 2013; Harra et al., 2007; Sterling & Hudson, 1997). The relationships between dimming areas, flare energy release, and the ejection of magnetic flux ropes as parts of a CME are complex and often inconsistent. Most of the above discussions emphasize the structure of the hottest parts of the flare such as the sigmoid or anti-sigmoid shape of the hot channel in flares above the PILs seen in AIA-335 or AIA-193. We emphasize the images from AIA-171, which show the cooler structures in the low corona as well as the flaring, hottest areas. We identify the ejected magnetic flux ropes by their outward motion in these AIA-171 images then trace their structures back to the photosphere where we place the MOPs on the basis of the surface field strength and field signs. The case of MFI 129 is discussed below in section 3.2.1.1 as an illustration of these considerations.

Our goal in this study is to find a way to make event specific helicity predictions. At present helicity predictions rely on the hemispheric-scale systematics in the sense of helicity, which were pointed out by Bothmer and Schwenn (1998) and are known as the Bothmer-Schwenn rule. The recent study by Savani et al. (2015) explicitly used this rule for their prediction of the chirality of the magnetic clouds. While this rule had been anticipated by physical arguments (Seehafer, 1990), conditions in active regions are complex so that its validity is reduced by several mechanisms (Leamon et al., 2002; Wang, 2013). The Bothmer-Schwenn rule is not specific to any particular event so that only general predictions can be made with it.

The relationship between coronal structures and magnetic clouds is well studied and has been summarized by Crooker and Horbury (2006). The further connection to measured solar surface magnetic fields is less well studied and has been done in the context of coronal dimming and solar flares. Martin and McAllister (1997) found that the chirality of erupting filaments can be identified from the locations of barbs on the side of the bright filament. Savani et al. (2015) studied eight cases to make a connection to solar surface features and in most cases we are not able to confirm their identifications. Our primary goal is to use identified solar surface sources of the magnetic clouds to develop methods of predicting cloud helicity. The identification of the solar sources for the clouds is thus a prerequisite to our effort. However, the identification of solar surface origins of CMEs is complicated by the fact that the CMEs are identified on the LASCO c2 coronagraph images while the originating ejected loop or cloud is seen on the disk as a bright or dark filament on the AIA EUV images without necessarily being connected to the CME matter. Flares often accompany the CME ejection but not always. The identification of the photospheric positions of the ejected flux rope has relied on the positions of coronal dimming that results from the loss of material in the CME (Harra et al., 2007; Sterling & Hudson, 1997). While this phenomenon can help identify the occurrence of a CME, the areas with dimming do not move outward in the cases we have studied. We search for outward moving loops in the AIA EUV images whose legs can be traced back to points at the solar surface. We identify these moving loops as the initial stage of the CME ejection if the time of the motion is appropriate for the CME and if the direction of the motion is correct for the subsequent pattern of the CME. The next step is to connect these moving loops with the solar surface magnetic fields and identify the MOPs.

At an early stage of the loop ejection, the legs can be traced back to a point where they end near the solar surface. This is not sufficient to identify the MOPs however. In keeping with the sketch of Figure 1 we make the location of the MOPs more precise by imposing the condition that they are at extrema points of opposite radial magnetic field. We describe these steps in more detail in the next section.

## 2. Methods

There are two parts to our estimation of these cloud properties: the identification of the MOPs for each cloud and the determination of the solar surface magnetic field configuration at those points. These two parts are carried out simultaneously through iteration rather than being sequential steps. The solar surface magnetic field generally has its strongest component in the radial direction. However, the helicity comes mostly from the fields parallel to the solar surface which near disk center are largely transverse to the line of sight. These magnetic field components can be determined from vector magnetograph data or from the line of sight magnetic field using the change in projection angle due to solar rotation. We derive our transverse field estimates from the HMI data in two ways based on these approaches as described below in section 2.2 . In particular we use the the derived data series from Ulrich and Tran (2016) available as described in section 4.6. This series begins 2011/01/07 so that the study period is from 2011/01/07 to the end of the MFI catalogue at 2012/12/31.

The prediction task we face is to work backward from the magnetic cloud at 1 AU near Earth to find an appropriate CME from the CDAW catalogue provided by the LASCO team (https://cdaw.gsfc.nasa.gov/CME_list/),

then identify an event on the solar surface that could have produced the CME. The availability of full-disk magnetic field measurements from HMI on SDO with a full-time cadence of 720 s is a major resource for the analysis of the magnetic fields and their evolution. The identification of the paired emergent points with strongest, opposing radial magnetic fields for the ejected flux tube arch is then an essential part of the estimation of the flux-tube helicity.

We restrict this study to the 37 clouds listed in the MFI table after the beginning of 2011. This table gives starting and ending times for the detection of the cloud at 1 AU near earth along with values from the Lepping et al. (1990) model for the helicity and strength of the cloud magnetic field. The connection between solar surface events and observed magnetic clouds at 1 AU is not simple due to poorly known cloud velocities, uncertain direction of propagation for the clouds and interactions between the clouds and the intervening heliospheric magnetic field. Consequently, we have adopted a set of requirements given below that need to be met before we can identify the solar origins of the cloud and then estimate the cloud properties.

### 2.1. Methods for Identification of Solar Surface Events With MFI Clouds

We seek a connection between the observed magnetic clouds and a specific solar surface event. For the largest solar events such a connection may be obvious but for most of the clouds listed in the MFI table there are a number of candidates or in some cases, none. The critical consideration is the time lag between the solar event and the observation of the cloud at 1 AU. The cloud velocity $v_{cloud}$ at 1 AU is given in the MFI table along with start $t_{start}$ and end $t_{end}$ times for the cloud passage. We calculate a preliminary solar event time $t_\odot$ from

$$t_\odot = (t_{start} + t_{end})/2 - 2 * 1714/(v_{cloud} + v_{eject})$$ (1)

where the times are in days and the velocity is in km/s, we start with $v_{eject} = 500$ km/s as a typical speed for a CME and the coefficient 1,714 converts a travel distance of 1 AU minus $2.2R_\odot$ into a travel time of days. We use the CDAW catalogue to find candidate halo or partial halo CME near estimated time of origin. For CMEs seen in LASCO C2 and C3 videos, the travel time takes into account the fact that the cloud emerges from a point $1.2R_\odot$ above the photosphere. If there is a halo or partial halo CME within 12 hr of this time we consider it to be a candidate and examine the position and direction of CME travel using the C2 and C3 videos. The CDAW catalogue gives three estimates for the CME velocity and we replace the initial guess velocity with the average of those three to get a new ejection time which usually does not match the actual ejection time. In order to reconcile the derived ejection time with the actual ejection time and thus validate the candidate halo or partial halo CMEs we allowed for an adjustment of the ejection velocity by as much as $\pm 600$ km/s subject to the condition that $v_{eject}$ remains above the $v_{cloud}$. In case there are multiple candidates we choose the one with the smallest velocity adjustment.

If a candidate CME appears to have been ejected toward Earth then we examined it further using the jHelioviewer (Müller et al., 2017): http://www.JHelioviewer.org (we refer to this as the jHv tool). The jHv tool allows for the use of many types of image and allows for these to be displayed as observed or as either running differences or as fixed base differences. In accordance with our strategy we look for a loop structure that moves outward from the solar surface travelling from the quadrant identified in the C2 and C3 videos and moving in the direction indicated for the CME motion. The start time for the outward motion needs to be 15 min or less before of the initial appearance of the CME. We use the AIA-171 to carry out this search. The cadence of the images we used was 5 min. A shorter interval seemed to be less effective in identifying outgoing flux ropes while for the examination of dimming areas a cadence of 2 min was used. For context we searched beginning 120 min before the expected CME. The running difference images were the definitive resource for the search. The base difference approach shows the coronal dimming but these images appear largely unrelated to the outward moving flux ropes.

### 2.2. Estimation of Solar Surface Magnetic Field at the Origination Points

We derive our transverse field estimates from the HMI data in two ways: using line-of-sight fields along with the variable projection angles due to solar rotation as given by Ulrich and Boyden (2006) and extended by Ulrich and Tran (2016; Ulrich-Boyden-Tran, UBT) and from the Solar Software (SSW) package available at http://www.lmsal.com/solarsoft/ssw_install_howto.html with further instructions on use at https://www.lmsal.com/sdodocs/doc/dcur/SDOD0060.zip/zip/entry/. Both the UBT and SSW approaches use a procedure called derot*mean*, which takes an average over a specified time interval with each data value shifted in longitude or central meridian angle (CMA) to the position where the observation would be at the specified time for the output map. The time changes applied in adding a new image to the building up average image can be either positive or negative with the final image typically but not necessarily being at the center time for the interval. All magnetic field components can be handled this way. The interval we use for the input to UBT and SSW analyses is 1 or 2 hr. The regions of interest are near the central meridian and the time interval is short so that rotation does not impact the visibility of the point.

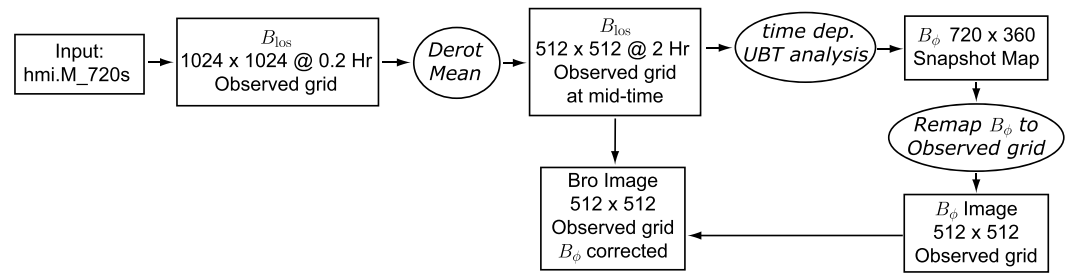

**Figure 2.** This chart gives the flow of data through the steps leading to the products we use. The rectangular boxes denote the data products, not all of which are saved, while the ovals give the processing steps between the products. The spatial format is either in an as-viewed perspective format, which we refer to as the *observed grid*, or in a synchronic map format, which we refer to as a *snapshot map*. The maps have quantities evaluated over the full solar surface including those portions not visible at the mapping time. These values are derived as part of the UBT method, which is essentially a time-dependent least-squares fit linear interpolation algorithm. The boxes indicate data set sizes and format as well as the temporal cadence. The Bro quantity combines $B_{los}$ from derot mean corrected for projection effects and corrects for the transverse component using the UBT estimate of $B_\phi$. This flow chart lists the observed grid as having a $512 \times 512$ format, which was used after the magnetic origination point positions were determined. The initial step to identify the solar surface sources used images in the $256 \times 256$ format. UBT = Ulrich-Boyden-Tran.

The UBT and SSW approaches differ in two important ways: (1) The SSW method uses high temporal and spatial resolution images getting the transverse fields from vector magnetogram data, while the UBT method uses coarser temporal and spatial resolution images but does not require vector magnetogram data; (2) the SSW data depends on conditions at the time of the event, while the UBT approach includes dependency on conditions over the prior Carrington rotation. We see below that in at least one case this dependency on the prior rotation may be an advantage. Also, the lack of a need for vector magnetogram data permits the application of the UBT method to times going back to the year 1986.

### 2.2.1. The UBT Estimation

The line-of-sight (los) magnetic field as measured by the longitudinal Zeeman splitting has lower noise than does the transverse magnetic field which is derived from Stokes polarimetry. Measurements of the los field are also available for a larger area of the solar disk than is the case for the vector field. Furthermore the archival data sets generally have a greater representation of the los field than the vector field. A method of obtaining an estimate of the component of the field in an east/west direction was used by Shrauner and Scherrer (1994) and Ulrich and Boyden (2006). The basic idea from these two publications is that the line of sight field, $B_{los}$ includes a contribution from the radial field $B_r$ of the form $B_r \cos(CMA)$ and a contribution from the E/W field $B_\phi$ of the form $B_\phi \sin(CMA)$. Solar rotation carries points on the solar surface past our line of sight so that the CMA changes with time. Those methods could not distinguish between variation due to time dependence of the radial field and variation due to the changing CMA. We use a method for incorporating both time dependence and changing CMA which we call the UBT analysis.

Differential rotation causes solar features to be mapped at locations on a full spherical coordinate system which are different from the observed locations when the mapping time $t_{map}$ is different from the observed time $t_{obs}$. Consequently, a map of the solar surface $B_r$, for example, must specify $t_{map}$. This location shift can be done when there are large differences between $t_{obs}$ and $t_{map}$ although in such cases evolutionary changes must also be taken into account. The essential idea of the UBT approach is that by going back in time by one rotation period, the projection effects are the same, so the observed change is due to temporal variations of the magnetic field components. (Ulrich & Tran, 2013, 2016) explained the process of generating full-surface maps showing quantities at a single time, $t_{map}$. They called these snapshot maps. The snapshot maps are now also referred to as synchronic maps. The steps required to generate the snapshot maps with UBT approach are shown in a flow chart format in Figure 2.

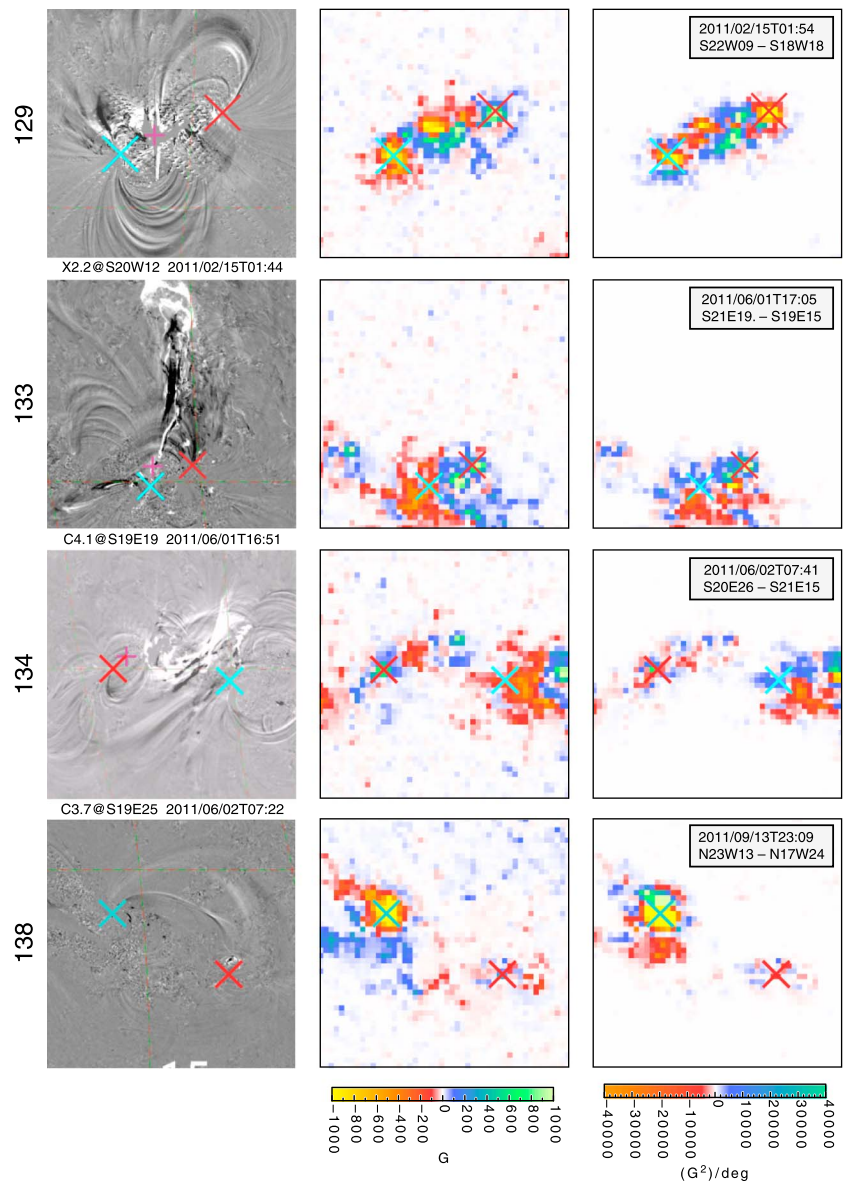

**Figure 3.** Properties of the first four identified events. Each set of three maps applies to one of the Magnetic Field Investigation entries whose numerical index is given to the left of the row. The leftmost column gives the jHv image with the locations of the magnetic origination points shown with the colored "X" marks. The colors are opposite to the colors indicating the Bro polarities. Flare properties: peak emission, time, and location are given below each jHv image. The flare location is shown by the purple "+" sign. The middle column gives the Bro map, and the right column gives the *IH* map. The locations of the magnetic origination points are given in the upper corner of the plot. The plotted locations are based on the higher-precision positions given in Table 1. Scales for both Bro and *IH* are given below the right two columns. The size of both sides of the crop box is $R_\odot/2.7$.

The intermediate steps involve transformations between a variety of map formats that can be used to display and process the 2-dimensional data. Observations of the solar disk obtained by HMI and AIA are on what is called helioprojective coordinates (Thompson, 2006), which we refer to as the *observed grid*. The UBT analysis, which is essentially a least-squares-fit interpolation process, works in the snapshot map format and collects all the observations of the los magnetic field within a specified time interval and for which each point has $|CMA| \leq 75°$. We multiply each magnetic field observation by geometric and time-dependent factors to obtain a set of four equations for $B_r$, and $B_\phi$ and their time derivatives. For this processing we use a snapshot map format with a grid of 720 by 360 in longitude and latitude, respectively. Details of the method are given in Ulrich and Boyden (2006) and (Ulrich & Tran, 2013, 2016).

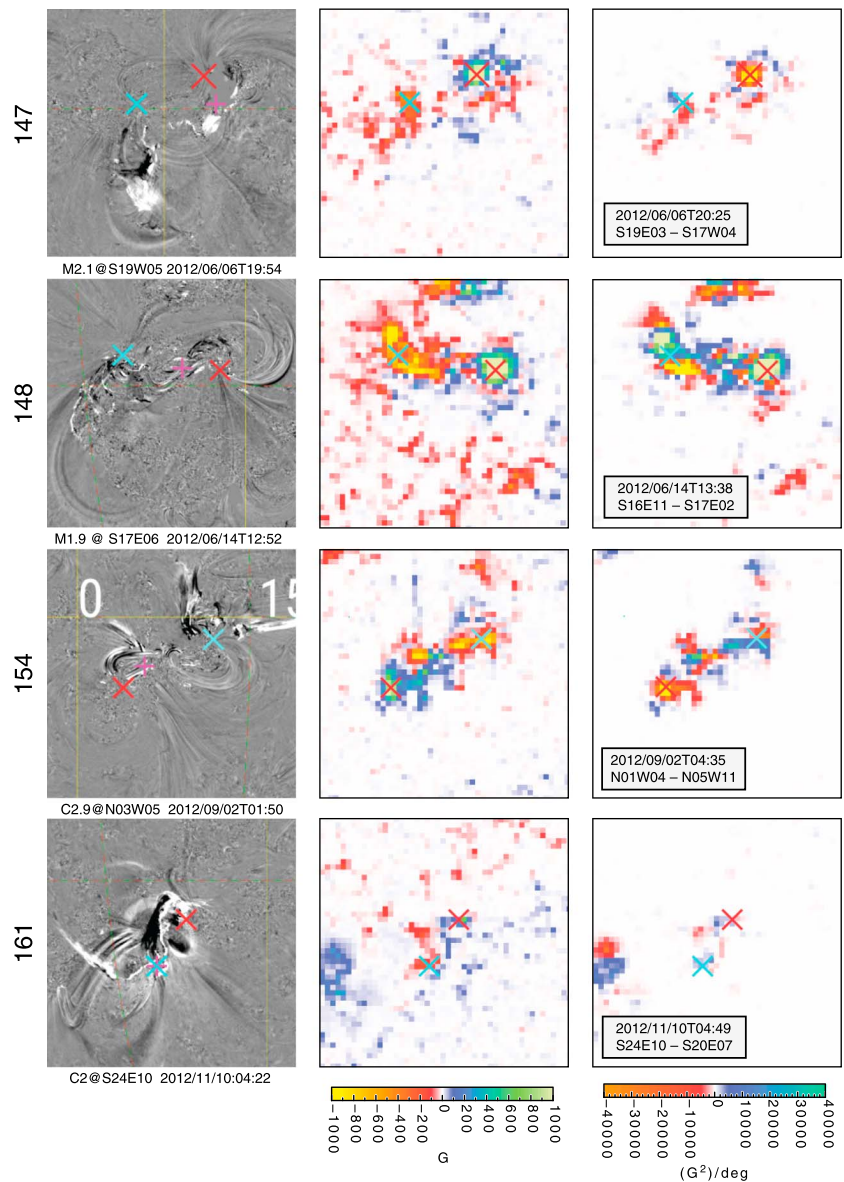

**Figure 4.** Properties of the second four identified events labeled as in Figure 3.

Typically in our application the time interval goes from the mapping time back about one Carrington rotation period and forward beyond the mapping time until the point exits the west limb of the solar image. This procedure uses observations after the mapping time which is possible for a retrospective analysis but not for the case where the mapping time is the present time. Although the observations could be added individually, we use the 2-hr average images from the derot*mean* procedure to shorten the computing time needed for the snapshot maps. We then project $B_\phi$ at the mapping time onto either a $256 \times 256$ or a $512 \times 512$ observed grid map. On the observed disk we also have the line-of-sight magnetic field at the mapping time which is not dependent on any interpolation process except the derot*mean* averaging. In order to use magnetic field values that are most correct for the time of the map, we adopt an approach that combines the line-of-sight field at the mapping time with a correction for the east-west field from the UBT analysis yielding a hybrid quantity we call Bro. Thus, the two quantities we use are $B_\phi$ and Bro on the observed grid. We work with these images and the jHv images, which are also in the observed grid format to determine the MOP locations and to estimate the helicity.

**Table 1**
*Properties of Identified Magnetic Cloud Sources*

| MFI | | 1 AU cloud start time | | | | Solar source | Handedness | | Magnetic origination points | | | | Sep | MOP field | | B_o | MOP helicity | |
| --- | --- | --- | --- | --- | --- | --- | --- | --- | --- | --- | --- | --- | --- | --- | --- | --- | --- | --- |
| No. | Active region Number | Yr | Mon | Day | Hr | Yr/Mo/Day:Hr:Min | Pred | Obs | CMA1 | lat1 | CMA2 | lat2 | | Bro1 | Bro2 | | IH1 | IH2 |
| 129 | 11158 | 11 | 2 | 18 | 20 | 11/02/15:02:24 | L | L | 8.83 | −21.85 | 17.97 | −17.98 | 9.33 | −860 | 1200 | 11.4 | −77000 | −150000 |
| 133 | 11226 | 11 | 6 | 5 | 1 | 11/06/01:18:36 | R | R | −19.40 | −20.83 | −15.09 | −18.97 | 4.41 | 26 | 624 | 20.5 | 4400 | 37000 |
| 134 | 11226–11227 | 11 | 6 | 5 | 9 | 11/06/02:07:41 | R | R | −26.32 | −20.31 | −14.70 | −21.27 | 10.90 | 620 | −215 | 8.6 | 26000 | 7300 |
| 138 | 11289–11293 | 11 | 9 | 17 | 16 | 11/09/13:23:12 | L | L | 13.18 | 23.10 | 23.57 | 17.39 | 11.12 | −2000 | 200 | 17.2 | −750000 | −4300 |
| 147 | 11494 | 12 | 6 | 6 | 20 | 12/06/06:20:36 | L | L | −2.47 | −18.99 | 3.51 | −16.51 | 6.17 | −360 | 1600 | 9.9 | −2900 | −13500 |
| 148 | 11504 | 12 | 6 | 16 | 23 | 12/06/14:14:12 | R | R | −10.69 | −15.96 | −2.06 | −17.25 | 8.36 | −620 | 1700 | 36.3 | 27000 | 280000 |
| 154 | 11560–11561 | 12 | 9 | 5 | 6 | 12/09/02:04:00 | L | L | 3.55 | 1.19 | 11.36 | 5.20 | 8.77 | 876 | −775 | 14.2 | −84000 | −4800 |
| 161 | 11610 | 12 | 11 | 13 | 9 | 12/11/10:05:12 | R | R | −10.41 | −24.21 | −7.35 | −19.83 | 4.95 | −140 | 290 | 25.8 | 2000 | 1100 |

1 AU cloud start time: The MFI table also gives end times for the magnetic clouds.
Solar source: The time of the beginning of the outward motion of the CME.
Handedness: The predicted and observed helicity handedness.
Magnetic origination points: Angular heliographic positions of the MOPs.
Sep: Angular separation of the two MOPs.
MOP field: The magnetic field strength $Bro$ in gauss as found by the Ulrich-Boyden-Tran analysis.
$B_o$: The cloud model field strength given in the MFI table in nT.
IH: The helicity index giving the value of $B$, times $(\nabla \times B)$, in units of $G^2$/degree.

**Table 2**
*Reasons for the Failure to Make an Association With a Solar Surface Event*

| MFI No. | Reason | | | | |
|---|---|---|---|---|---|
| | A | B | C | D | E |
| 127 | X | X | | | |
| 128 | | | X | | |
| 130 | | | X | | |
| 131 | X | | | | |
| 132 | X | | | X | |
| 135 | | | X | | |
| 136 | | | X | X | |
| 137 | | X | | | |
| 139 | | X | | X | |
| 140 | | | X | X | |
| 141 | | X | | X | |
| 142 | | | X | X | |
| 143 | | X | | | |
| 144 | | X | | X | |
| 145 | | | | X | |
| 146 | | | | X | |
| 149 | | | X | | |
| 150 | | | X | | |
| 151 | X | | | | |
| 152 | X | | | | |
| 153 | X | X | | | |
| 155 | | | | | X |
| 156 | | X | | | |
| 157 | X | X | | | |
| 158 | | X | | X | |
| 159 | | | | X | |
| 160 | | X | X | | |
| 162 | | | X | | |
| 163 | X | | | | |

*Note.* MFI = Magnetic Field Investigation.
A: No strong CDAW event.
B: Weak activity.
C: Too far away from disk center.
D: Timing too far-off.
E: Magnetic region too complex.

### 2.2.2. Using the SSW Package

The second method we use relies on the SSW package of reduction software. The full vector magnetogram data from the HMI system (Hoeksema et al., 2014) as a function of heliographic coordinates is available in the Spaceweather HMI Active Region Patches (SHARPs; Bobra et al., 2014). The primary drawbacks for the SSW approach are the relatively high noise in areas of low magnetic field and the need to resolve the 180° ambiguity in the field direction. The ambiguity resolution in the SSW package is done by the method given by Leka et al. (2009).

### 2.2.3. Estimating Helicity

Magnetic helicity is a volume integral of the dot product of the magnetic vector potential and the magnetic field. Depending on the geometry in question there can be contributions from both twist (rotation of the flux tube in a plane perpendicular to the field direction) and writhe (a large scale kink in the 3-dim structure of the flux tube; Blackman, 2015). The observations available to us are at the solar surface and are essentially two-dimensional so we focus on the twist. At a point associated with a peak magnetic field as indicated in Figure 1 the twist is proportional to $\nabla \times \mathbf{B} = \frac{4\pi}{c}\mathbf{J}$ where $\mathbf{J}$ is the local current density. To estimate the handedness of the magnetic region we use $(\nabla \times \mathbf{B})_r$ times $B_r$:

$$IH = B_r\ (\nabla \times \mathbf{B})_r \tag{2}$$

which will be positive for a RH flux tube and negative for a LH flux tube. *IH* is proportional to the local current helicity density. The quantity $(\nabla \times \mathbf{B})_r$ involves two angular derivatives: $\partial B_\phi/\partial\theta$ and $\partial B_\theta/\partial\phi$. For the evaluation of the helicity using the UBT approach, only $\partial B_\phi/\partial\theta$ is available so in fact, we cannot calculate the local current helicity density. Because the helicity of the field is a consequence of solar differential rotation which produces shear in the field, we expect that the helicity will have this derivative as a major contributor and we recognize that the numerical values we find for $B_r * (\partial B_\phi/\partial\theta)$ are an index with somewhat arbitrary values which should indicate the relative strength of the twist. Pevtsov and Latushko (2000) used a similar technique to estimate the helicity and pointed out that the derivative components we have omitted average to zero. By comparison the SSW approach uses vector magnetograph data which gives both $B_\theta$ and $B_\phi$ along with numerical derivatives needed to get *IH*. We show comparisons between the two methods in Figures 3 and 4 which show that there are both similarities and differences between the two approaches.

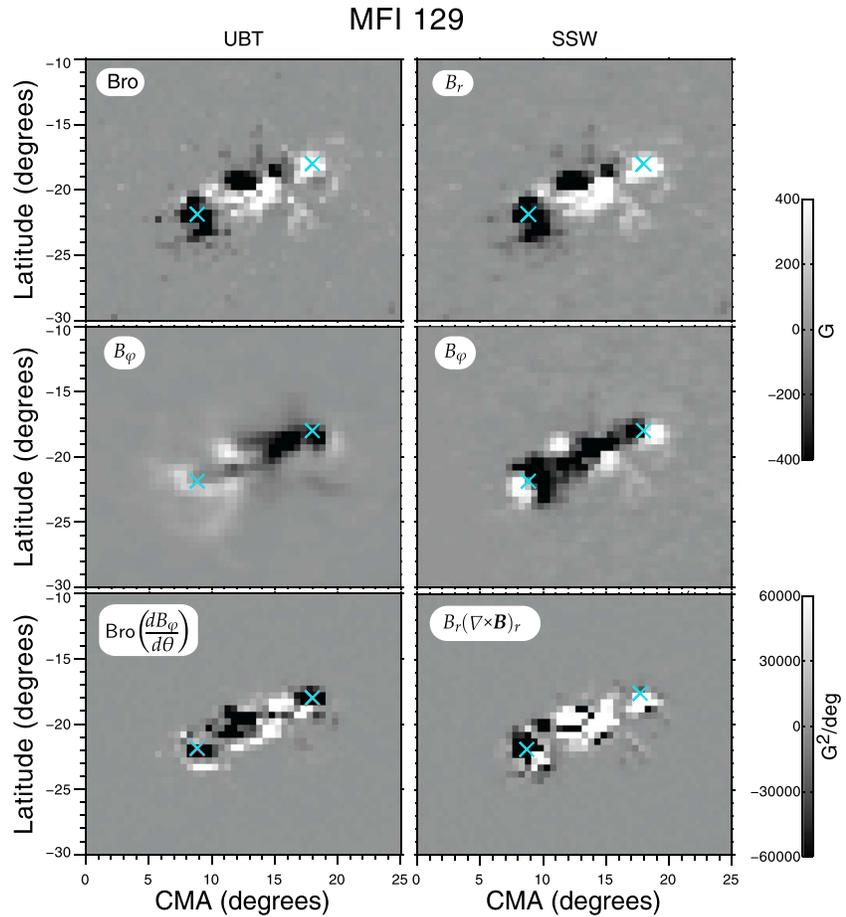

**Figure 5.** Comparison between the results from UBT analysis and the SSW results. The spatial resolution for the SSW case has been degraded to match the of the UBT analysis. The plotted quantities differ but are the closest to each other that are available from the different reduction methods. MFI = Magnetic Field Investigation; SSW = Solar Software; UBT = Ulrich-Boyden-Tran.

## 3. Solar Surface Events Associated With MFI Clouds
### 3.1. The Associated Events
Using the procedures given above we were able to associate 8 of the 37 entries in the MFI table with solar surface events leaving 29 nonassociated entries. We give summary properties of the 8 associated events in

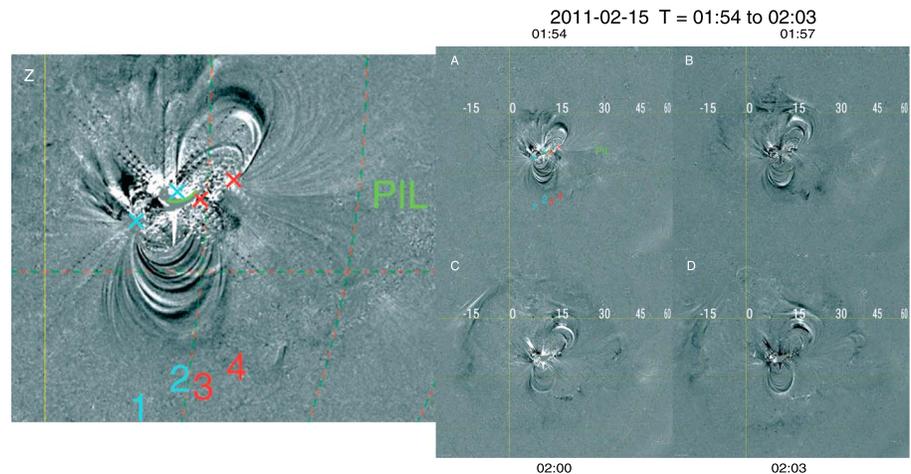

**Figure 6.** A sequence of frames from the jHv tool showing running difference images from AIA-171 for the loop ejection for MFI 129. The starting frame (a) also includes designations of points of extrema in the radial field strength. A zoomed image of these designations is given on the left in panel (z). The numbers below the X marks label the points without overwriting any of the important features. Points 1 and 4 are the ones identified in this section as the magnetic origination points.

Table 1 and indicate reasons why we failed to make an association in Table 2. Timing consistency indicates that we were able to find a velocity adjustment that brought the CDAW CME ejection time into agreement with the cloud observation time using the simple formula of Equation 1. The identifying images and the magnetic data leading to the properties listed in Table 1 are given in Figures 3 and 4. The quantities shown in these figures are the jHelioviewer running difference image, the Bro image and $IH$ image. The jHv images are selected at 4-min intervals and the figures show structure at the time we judge to be closest to the ejection time. The magnetic maps were calculated at intervals of 2 hr and the ones shown are at the time closest to that of the ejection. The colored X marks are at the locations of the MOPs (corrected for rotation between the jHv and magnetic map times). We carried out a search of the X-ray record on the CDAW site after the identifications were made to find associated flares which occurred close to the times of the CME ejection. Flares were found for seven of the eight events. Properties of these flares are given in Figures 3 and 4. The flare location is shown on the jHv figures using a purple "+" sign. We also compare the results from the UBT approach to that from the SSW system in Figures 5 and 8. The first case is for MFI 129 where the two approaches give different predictions for helicity, and the second is for MFI 138 for which there is good agreement. The case of MFI 129 is the only instance where the SSW method gave helicity that disagrees with the observed handedness of the cloud at 1 AU.

Two previous publications have included events within our study period and have provided explicit connections between solar surface events and magnetic clouds near earth: Li et al. (2014) and Savani et al. (2015). Due to the number of CMEs observed leaving the sun as halo or partial-halo events and our flexibility with the travel time condition, there is considerable ambiguity in the connections. We refer to those identification by Li et al. (2014) as the Li events and those by Savani et al. (2015) as the Savani events. For the Savani events we reference their event ID as Savani-a to Savani-g.

### 3.2. Properties of Identified Events
### 3.2.1. MFI 129
The CDAW record shows a clear halo for the identified time. The travel time is too short if the average of start and end time for the magnetic cloud are used. However, if only the start time is used then the ejection time agrees with the observed CME time with only a −85-km/s ejection velocity adjustment. With the average of the starting and stopping times the required adjustment is −230 km/s. The event duration, stop time minus start time, is larger than most and this contributes to the required adjustment velocity.

Two pairs of loops are launched at the same time. This event was highlighted by Hoeksema et al. (2014) in part because of the class X flare at the time of the CME eruption. The SSW analysis gives the opposite helicity from the UBT analysis for this case. The transverse field $B_\phi$ derived from the UBT analysis depends on the field strength of both the current and the previous Carrington rotations so that evolutionary factors can influence the derived values. It is evident from the comparison in Figure 5 that there are differences in the results of the two methods.

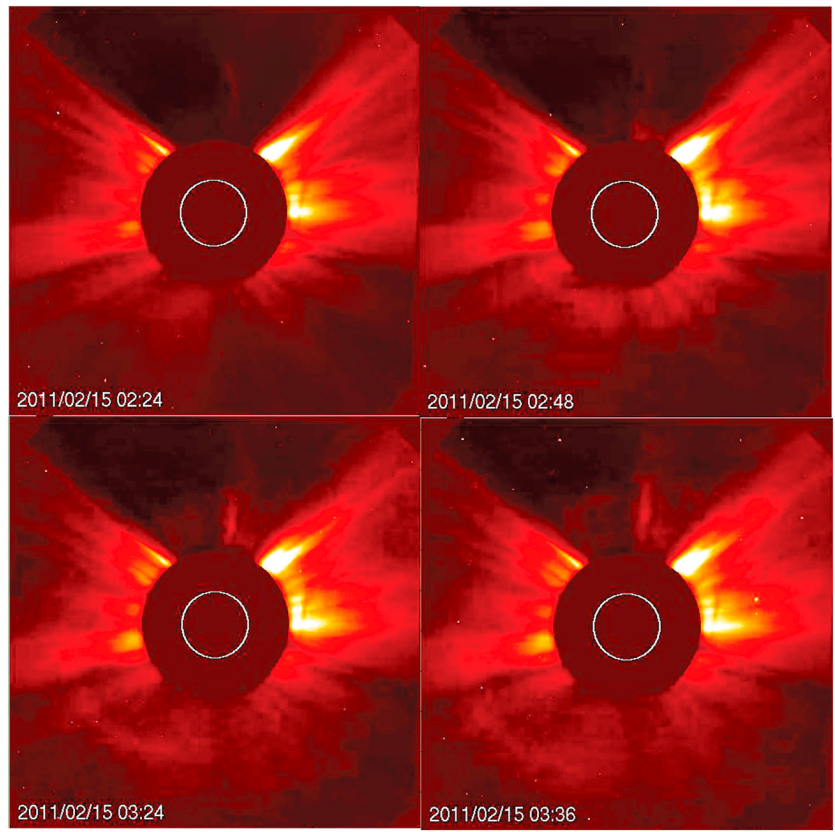

**Figure 7.** Images from the Large Angle and Spectrometric Coronagraph C2 video sequence showing the emergence of the coronal mass ejection for Magnetic Field Investigation 129 beginning at about 02:40.

The evolution of this active region was quite complex with shear and strong flares. Because this event included the first strong flare of the solar cycle and because the SDO/HMI and SDO/AIA systems had just become available there is an extensive literature analyzing many aspects of the active region and its magnetic fields. Tziotziou et al. (2013) have provided a concise but thorough review of the publications discussing various aspects of AR 11158. They also analyze the total helicity for the region with a discussion of the relationship between the total helicity and the triggering of flares. Song et al. (2013) investigated the deviations from potentiality for the magnetic field configurations. One of their indicators is the fractional current helicity for the whole active region. They found the positive current helicity to exceed the negative current helicity by about a factor of two. This agrees with our determination using the SSW method. The magnetic structure and its relationship with the flare were discussed by Zhao et al. (2014). At the time of the flare the active region included a hyperbolic flux tube with a quadrapole magnetic field configuration.

The structure for this case in Figure (3) shows that the ejected loop has the dipole pattern we assume. The primary cause for the difference between the UBT and SSW prediction is the configuration of $B_\phi$ at the westward MOP. Given the structures of $B_\phi$ and $IH$ shown in Figure 5 for the UBT approach, the prediction of LH for the chirality is clear. However, there are many details in the structure and evolution of the magnetic field of the region that are not part of our assumed model.

### 3.2.1.1. Identification of the Outwardly Moving Magnetic Flux Rope

The CME leading to MFI 129 observed at 1 AU has been discussed in detail in a number of publications as listed above. The work by Janvier et al. (2014) includes detailed maps of the vertical current density finding that the pattern of the strongest of these currents has a *J* shape with the hook of the *J* surrounding areas of coronal dimming. They identify these regions as the foot points for the flux rope ejected by the CME and give a sketch of the configuration they find. Their study is based on observations of high temperature regions as shown by AIA-335 images. The features shown in the AIA-335 images are clearly the sources of the energy release in the flare but the flux rope shown in their Figure 7 sketch is not seen in the AIA-171 images, which are sensitive to the cooler matter that is not part of the primary energy release. We used the jHv tool to extract four running difference images from the AIA-171 images at times near the primary energy release for the flare using a time step of 3 min. These images are shown in Figure 6 as panels (a) to (d). The grid lines and angle values on these maps are relative to the image center and correspond to CMA and the difference between latitude and the latitude of the disk center $b_0$. The numbers in Figure 6 show the locations of the magnetic field extrema. Points 2 and 3 correspond to the foot points identified by Janvier et al. (2014), while points 1 and 4 are the MOPs we have identified here. Figure 7 from Janvier et al. (2014) shows points 2 and 3 connected by the magnetic flux tube whereas we see from panel (a) that arches connect points 1 and 3 as well as points 2 and 4 but not points 2 and 3. Panels (b) through (d) show an outgoing wave of darkening that we interpret as the matter in the ejected CME. The loops or arches show changes that also appear to be outgoing motion but it is not clear whether this is due to interaction with the outgoing matter or actual outward motion since the structure of the loops and arches remains largely unchanged after the passage of the wave of darkening. The wave of darkening moves about $0.25R_{\odot}$ in 5 min corresponding to a speed of about 600 km/s. Figure 7 shows that the CME emerges at about 02:40 or about 37 min after the last frame of Figure 6. This lag is roughly consistent with the speed of the wave of darkening.

The connection between the wave of darkening and foot point or MOPs is not definitively clear. Neither the loops going from points 1 to 3 nor the loops going from point 2 to 4 are exclusively part of the outgoing wave of darkening. Parts of the loops seem to join the wave, but in fact, the wave appears more like a blast wave ejecting matter already far from the flare center. We have assigned points 1 and 4 to be the MOPs because these are the strongest field points and they have likely dominated the preexisting magnetic configuration. The UBT analysis may give the correct prediction of the cloud helicity because it incorporates effects from the magnetic configuration prevailing over the previous rotation.

### 3.2.2. MFI 133

This is a complex region with an eruption from the W side of the group. During the early stage the erupting loop has its W side in an R region and its E side in an L region. The W side has stronger helicity and during the peak eruption the E side moves into the clearly R region. The velocities in the CDAW catalogue are less than the cloud velocity. Our identification increases the ejection velocity so that it equals the cloud velocity in order to have consistent timing. Li et al. (2014) picked the time of the cloud to be 2UT instead of 5.3 UT as given in the MFI table. They then identified a flare 14 hr later at S18E28 as the source for this cloud. With the MFI table timing the ejection velocity has to reduced 250 km/s. The source points then have to be assigned to the magnetized area to the east of those shown in Figure 3, and the predicted helicity then becomes L instead of R.

### 3.2.3. MFI 134

Two eruptions are found in the jHv images separated by just over 1 hr. This is the second. The magnetic structure is rather complex. The loop and its foot points were chosen because the foot points go between opposite polarities and the loop is one of the structures initiating the eruption. The $v_{ejection}$ was adjusted $-440$ km/s to match the observed CME time. The second eruption was chosen because its higher ejection velocity means it would overtake and dominate the first ejection.

### 3.2.4. MFI 138

Several partial halo CME's occur on this date with the lag time needed for this this CME requiring no ejection velocity adjustment. There are two sets of loops which are both moving and undergoing intensity changes with a center magnetic area connecting both east and west. The loop ejected toward Earth connects the western two areas. The easterly loop does not have a clear filament ejection event but does flare. No ejection velocity adjustment needed. Our identification agrees with the Li identification.

### 3.2.5. MFI 147

This partial halo requires decreasing the ejection velocity by 200 km/s. The CME is in the southern hemisphere along the central meridian line. There is little loop motion or brightness change in the AIA 171 images associated with the strongest fields of the spot group. The clear ejection comes from a very weak field close to the trailing/eastern spot. Li et al. (2014) describe the surface activity as uncertain.

### 3.2.6. MFI 148

This is Savani-d. Matching the lag time according to equation (1) requires the ejection speed to be reduced by 150 km/s. Our identifications agree with those of Li et al. (2014) except that our origination points are a bit to the east of theirs.

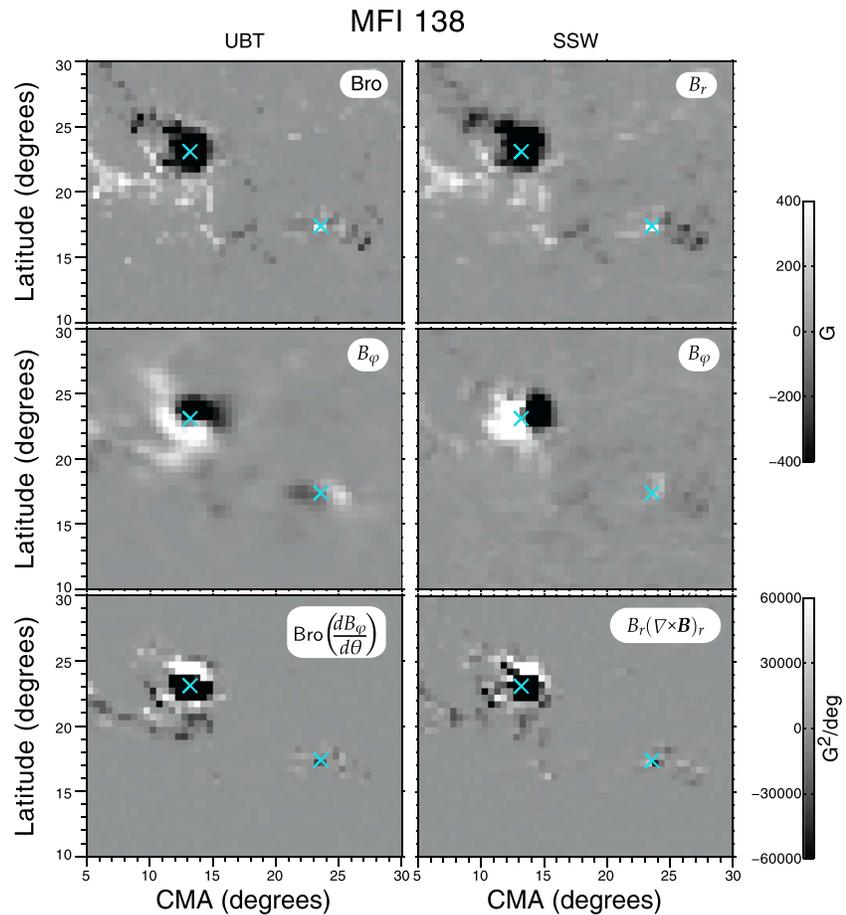

**Figure 8.** Comparison between the UBT analysis and the SSW results for MFI 138. MFI = Magnetic Field Investigation; SSW = Solar Software; UBT = Ulrich-Boyden-Tran.

### 3.3. The Savani Cases

A study by Savani et al. (2015) provides a number of identifications of CME events along with a discussion of these cases. The first case and the last two cases are outside our study period as defined above in section 2. The cases within our study period are Savani-b to Savani-f. Case Savani-b does not have a halo or partial halo associated with it in the CDAW catalogue. There are two small flares that could be associated with the cloud but there is no clear ejection of a loop except possibly during an eclipse period for SDO. The magnetic patterns near the possible ejected loop are complex with small areas of magnetic polarity and helicity. We are unable to predict the helicity for this case. Case Savani-c does not have an entry in the MFI catalogue that could be associated with the time and place given by Savani. Case Savani-d is identified with MFI 148 and was discussed above. Savani-e is not considered for the same reasons as Savani-c. Case f is left out due to the absence of clear loop motions. The main loop undergoes brightening and dimming but does not move outward. Were this case to be included, it would have given a prediction of helicity sign in agreement with observations. The loop shape is inverse S-shaped and the MOP helicity is negative but in a complex region.

## 4. Discussion

### 4.1. Context of the Investigation

This study attempts to relate magnetic clouds to solar surface events in order to explain the observed helicity of the clouds. Three things need to be connected: magnetic clouds in the table, CMEs in the CDAW catalogue and changing features on the solar surface. The only way to make progress without carrying out extensive numerical modelling of the precursor periods near the observed CMEs was to make the assumption that there have to be valid MOPs. This assumption is discussed in section 4.5 below. Because the definition of the MOPs is relatively simple and unambiguous, this step might prove useful for predictions in an operational context.

### 4.2. Relationship to Bothmer-Schwenn Rule

The Bothmer-Schwenn rule states that CMEs originating in the southern hemisphere are right-handed, while those originating in the northern hemisphere are left-handed. One can question whether the gain from the method developed here justifies the effort. Of the eight or nine events studied here, all have the handedness correctly identified. This may be a statistical fluctuation which can be checked by studying an additional set of events. Out of this set of events, two do not conform to the B-S rule: MFI-129 and MFI-147. Although the SSW and UBT approaches give differing predictions for MFI-129 they are in agreement for MFI-147 and the map like that shown in Figure 8 is in good agreement for the helicity prediction. This rate of agreement with the B-S rule is in accordance with the findings of Wang (2013) and Leamon et al. (2002).

### 4.3. Relationship to Flares

Following the recommendation 25 years ago by Gosling (1993) we have made the solar surface event identifications without requiring that a flare occur within or near the temporal window defined by the cloud and CME ejection speed. After the acceptance or rejection of the possible event we went back to determine the influence of the flare (if there was one) on the development of the loop ejection. In most cases the flare produced a prompt thermal response in the form of loop brightening then dimming. This is easy to interpret as loop motion, but such an interpretation may or may not be correct. We attempted to resolve this uncertainty by comparing the loop position before and after the brightening period which was usually short compared to the overall ejection event. If the loop was present in a stable location before and after the brightening period the loop was not considered as the source of the CME. The case of Savani-f included a faint and rapidly moving pattern that was not a loop but was probably material ejected directly from the flare. The filament connecting the candidate MOPs, which had the reverse S pattern did not appear to be ejected.

### 4.4. Relationship Between Magnetic Clouds and Solar Surface Events

This work has highlighted the difficulty of associating specific magnetic clouds with specific solar surface events. Prior investigations have addressed some of these difficulties. The study by Marubashi et al. (2017) looked at 11 events preselected by the MiniMax24 Working Group 4 of the International Study of Earth-affecting Solar Transients (ISEST) and was able to identify only nine of those. Li et al. (2014) listed 24 clouds during our study interval and were not able to identify solar sources for just five. Their identifications included 14 on the MFI list of which we also identified eight. The case studies by Harrison et al. (2012) showed the helpful character of detailed propagation modeling to understand the association between solar events and magnetic clouds at 1 AU. Our study presented here differs from these prior studies in that we make use of the surface magnetic field configuration at the identified points to predict the helicity of the resulting ejected loops and ultimately in the magnetic cloud at 1 AU.

Our candidate list included 37 events in the MFI table and we were able to identify only eight (or nine if we include Savani-f). The occurrence of a loop or arch clearly moving outward was lacking for most of the cases left out. The range of time available for the event was extended by allowing the ejection velocity to be altered by as much as $\pm 600$ km/s, but, for some cases not among the eight discussed in this study, this flexibility was often not enough to include an acceptable candidate event. In an operational setting this algorithm would be applied to predict the helicity sign of an ejected cloud and these difficulties would not apply. Instead the uncertainty would apply to the probability that the cloud would encounter a particular part of the heliosphere like Earth or a spacecraft. The identification of the MOPs for a CME of interest remains as a task that for this study was done by a human investigator rather than by an automated algorithm. Removing this deficiency will require some development effort.

### 4.5. Validity of the Assumptions

The essential assumption for the helicity prediction is that the ejected material is related to the magnetic field structure at the photosphere and that the strongest radial magnetic fields dominate this relationship. The jHv running difference images show a range of structural changes at the time and place of the identified events. The linkage of these structural changes to the MOP location defined by the strongest field condition is not as clearcut as one would like. Some rapidly changing loops originate from areas with weak magnetic field. However, in the context of a magnetic cloud model with a helical field structure, the greatest number of field lines will be contributed by the areas of strongest field. We require two origination points of opposite polarity so that the flux tube can have an arch structure after it has left the solar surface regions. The complexity of the actual magnetic configuration makes it difficult to identify an ejected arch structure that can be traced back to the origination points. Our approach gets around this task by simply assuming that the origination points are at the positions of the strongest fields of opposite polarity. Whether or not this is a useful approach depends on its success in accounting for the observed helicity of magnetic clouds at 1 AU. At least for the clouds in the MFI table the method shows some promise.

### 4.6. Online Data

We have made available four years of the data derived from the UBT method with as-viewed format images on a 2-hr cadence. The directory containing these files is posted on an ftp server at ftp://howard.astro.ucla.edu/pub/obs/HMI.DerivedFD/512_2hr/, which has a subdirectory structure with names in a yyyy_mm format. The files within each subdirectory have names in the format: `Bro_x_dBphi_20120702_2200x2.fits` where the last 15 characters before the `.fits` give the year, month, day, time, and x2. The x2 indicates the file has 512 × 512 pixels instead of 256 × 256 pixels. Preceding the time identifier are a varying number of characters defining the quantity in the file. The quantities are as follows: Bro_x_dBphi—the current helicity density used in this paper *IH*, bphi—the transverse field in the $\phi$ direction, and bro—the hybrid value for the radial component of the magnetic field.

#### Acknowledgments

We thank Todd Hoeksema and Yang Liu for helpful comments on the manuscript. We thank the referees for their considerable guidance in improving this paper. Assistance from Sam Freeland was critical in getting the SSW package to work at UCLA. The HMI images used in preparation of this paper are courtesy of NASA/SDO and the HMI science team. This research has been supported by NASA through award NNX15AF39G to Predictive Science, Inc. and subaward to UCLA. Data used in this study are available at ftp://howard.astro.ucla.edu/pub/obs/HMI.DerivedFD/512_2hr/ and at http://jsoc.stanford.edu/new/#.

## References

Baker, D. N., Akasofu, S.-I., Baumjohann, W., Bieber, J. W., Fairfield, D. H., Hones, E. W., et al. (1984). Substorms in the magnetosphere. *NASA Reference Publication*, *1120*, 8–1.

Blackman, E. G. (2015). Magnetic helicity and large scale magnetic fields: A primer. *Social Science Research*, *188*, 59–91. https://doi.org/10.1007/s11214-014-0038-6

Bobra, M. G., Sun, X., Hoeksema, J. T., Turmon, M., Liu, Y., Hayashi, K., et al. (2014). The Helioseismic and Magnetic Imager (HMI) vector magnetic field pipeline: SHARPs—Space-weather HMI active region patches. *Solar Physics*, *289*, 3549–3578. https://doi.org/10.1007/s11207-014-0529-3

Bothmer, V., & Schwenn, R. (1998). The structure and origin of magnetic clouds in the solar wind. *Annales Geophysicae*, *16*, 1–24. https://doi.org/10.1007/s00585-997-0001-x

Burlaga, L. F. (2002). Review of magnetic clouds/flux ropes and types of ejecta. *American Astronomical Society Meeting Abstracts #200, Bulletin of the American Astronomical Society*, *34*, 752.

Burlaga, L. F., Behannon, K. W., & Klein, L. W. (1987). Compound streams, magnetic clouds, and major geomagnetic storms. *Journal of Geophysical Research*, *92*, 5725–5734. https://doi.org/10.1029/JA092iA06p05725

Carrington, R. C. (1859). Description of a singular appearance seen in the Sun on September 1, 1859. *Monthly Notices of the Royal Astronomical Society*, *20*, 13–15. https://doi.org/10.1093/mnras/20.1.13

Cheng, X., & Ding, M. D. (2016). The characteristics of the footpoints of solar magnetic flux ropes during eruptions. *The Astrophysical Journal Supplement Series*, *225*, 16. https://doi.org/10.3847/0067-0049/225/1/16

Cheng, X., Guo, Y., & Ding, M. (2017). Origin and Structures of Solar Eruptions I: Magnetic Flux Rope. *Science in China Earth Sciences*, *60*, 1383–1407. https://doi.org/10.1007/s11430-017-9074-6

Cheng, J. X., & Qiu, J. (2016). The nature of CME-flare-associated coronal dimming. *The Astrophysical Journal*, *825*, 37. https://doi.org/10.3847/0004-637X/825/1/37

Chertok, I. M., Grechnev, V. V., Belov, A. V., & Abunin, A. A. (2013). Magnetic flux of EUV arcade and dimming regions as a relevant parameter for early diagnostics of solar eruptions—Sources of non-recurrent geomagnetic storms and forbush decreases. *Solar Physics*, *282*, 175–199. https://doi.org/10.1007/s11207-012-0127-1

Committee on the Societal and Economic Impacts of Severe Space Weather Events: A Workshop, National Research Council (2008). *Severe Space Weather Events–Understanding Societal and Economic Impacts: A Workshop Report*. Washington, DC: The National Academies Press.

Crooker, N. U., & Horbury, T. S. (2006). Solar imprint on ICMEs, their magnetic connectivity, and heliospheric evolution. *Social Science Research*, *123*, 93–109. https://doi.org/10.1007/s11214-006-9014-0

Gonzalez, W. D., & Tsurutani, B. T. (1987). Criteria of interplanetary parameters causing intense magnetic storms (Dst of less than -100 nT). *Planetary and Space Science*, *35*, 1101–1109. https://doi.org/10.1016/0032-0633(87)90015-8

Gosling, J. T. (1993). The solar flare myth. *Journal of Geophysical Research*, *98*, 18,937–18,950. https://doi.org/10.1029/93JA01896

Hale, G. E. (1931). The spectrohelioscope and its work. Part III. Solar eruptions and their apparent terrestrial effects. *The Astrophysical Journal*, *73*, 379. https://doi.org/10.1086/143316

Harra, L. K., Hara, H., Imada, S., Young, P. R., Williams, D. R., Sterling, A. C., et al. (2007). Coronal dimming observed with Hinode: Outflows related to a coronal mass ejection. *Publications of the Astronomical Society of Japan*, *59*, S801–S806. https://doi.org/10.1093/pasj/59.sp3.S801

Harrison, R. A., Davies, J. A., Möstl, C., Liu, Y., Temmer, M., Bisi, M. M., et al. (2012). An analysis of the origin and propagation of the multiple coronal mass ejections of 2010 August 1. *The Astrophysical Journal*, *750*, 45. https://doi.org/10.1088/0004-637X/750/1/45

Hoeksema, J. T., Liu, Y., Hayashi, K., Sun, X., Schou, J., Couvidat, S., et al. (2014). The Helioseismic and Magnetic Imager (HMI) vector magnetic field pipeline: Overview and performance. *Solar Physics*, *289*, 3483–3530. https://doi.org/10.1007/s11207-014-0516-8

Hu, Q., Qiu, J., Dasgupta, B., Khare, A., & Webb, G. M. (2014). Structures of interplanetary magnetic flux ropes and comparison with their solar sources. *The Astrophysical Journal*, *793*, 53. https://doi.org/10.1088/0004-637X/793/1/53

Janvier, M., Aulanier, G., Bommier, V., Schmieder, B., Démoulin, P., & Pariat, E. (2014). Electric currents in flare ribbons: Observations and three-dimensional standard model. *The Astrophysical Journal*, *788*, 60. https://doi.org/10.1088/0004-637X/788/1/60

Klein, L. W., & Burlaga, L. F. (1982). Interplanetary magnetic clouds at 1 AU. *Journal of Geophysical Research*, *87*, 613–624. https://doi.org/10.1029/JA087iA02p00613

Leamon, R. J., Canfield, R. C., & Pevtsov, A. A. (2002). Properties of magnetic clouds and geomagnetic storms associated with eruption of coronal sigmoids. *Journal of Geophysical Research*, *107*, 1234. https://doi.org/10.1029/2001JA000313

Leka, K. D., Barnes, G., Crouch, A. D., Metcalf, T. R., Gary, G. A., Jing, J., & Liu, Y. (2009). Resolving the 180° ambiguity in solar vector magnetic field data: Evaluating the effects of noise, spatial resolution, and method assumptions. *Solar Physics*, *260*, 83–108. https://doi.org/10.1007/s11207-009-9440-8

Lemen, J. R., Title, A. M., Akin, D. J., Boerner, P. F., Chou, C., Drake, J. F., et al. (2012). The Atmospheric Imaging Assembly (AIA) on the Solar Dynamics Observatory (SDO). *Solar Physics*, *275*, 17–40. https://doi.org/10.1007/s11207-011-9776-8

Lepping, R. P., Burlaga, M. H. AcL. F., Farrell, W. M., Slavin, J. A., Schatten, K. H., Mariani, F., et al. (1995). The Wind Magnetic Field Investigation. *Social Science Research*, *71*, 207–229. https://doi.org/10.1007/BF00751330

Lepping, R. P., Burlaga, L. F., & Jones, J. A. (1990). Magnetic field structure of interplanetary magnetic clouds at 1 AU. *Journal of Geophysical Research*, *95*, 11,957–11,965. https://doi.org/10.1029/JA095iA08p11957

Li, Y., Luhmann, J. G., Lynch, B. J., & Kilpua, E. K. J. (2014). Magnetic clouds and origins in STEREO era. *Journal of Geophysical Research: Space Physics*, *119*, 3237–3246. https://doi.org/10.1002/2013JA019538

Linker, J. A., Mikić, Z., Riley, P., Lionello, R., & Odstrcil, D. (2003). Models of coronal mass ejections: A review with a look to the future. In M. Velli, R. Bruno, F. Malara, & B. Bucci (Eds.), *Solar Wind Ten, American Institute of Physics Conference Series* (Vol. 679, pp. 703–710). College Park, MD: American Institute of Physics. https://doi.org/10.1063/1.1618691

Martin, S. F., & McAllister, A. H. (1997). Predicting the sign of magnetic helicity in erupting filaments and coronal mass ejections. *Washington DC American Geophysical Union Geophysical Monograph Series*, *99*, 127–138. https://doi.org/10.1029/GM099p0127

Marubashi, K., Cho, K.-S., & Ishibashi, H. (2017). Interplanetary magnetic flux ropes as agents connecting solar eruptions and geomagnetic activities. *Solar Physics*, *292*, 189. https://doi.org/10.1007/s11207-017-1204-2

Müller, D., Nicula, B., Felix, S., Verstringe, F., Bourgoignie, B., Csillaghy, A., et al. (2017). JHelioviewer. Time-dependent 3D visualisation of solar and heliospheric data. *Astronomy & Astrophysics*, *606*, A10. https://doi.org/10.1051/0004-6361/201730893

Nitta, N. V., & Mulligan, T. (2017). Earth-affecting coronal mass ejections without obvious low coronal signatures. *Solar Physics*, *292*, 125. https://doi.org/10.1007/s11207-017-1147-7

Pevtsov, A. A., & Latushko, S. M. (2000). Current helicity of the large-scale photospheric magnetic field. *The Astrophysical Journal*, *528*, 999–1003. https://doi.org/10.1086/308227

Riley, P., Lionello, R., Mikić, Z., & Linker, J. (2008). Using global simulations to relate the three-part structure of coronal mass ejections to in situ signatures. *The Astrophysical Journal*, *672*, 1221–1227. https://doi.org/10.1086/523893

Savani, N. P., Vourlidas, A., Szabo, A., Mays, M. L., Richardson, I. G., Thompson, B. J., et al. (2015). Predicting the magnetic vectors within coronal mass ejections arriving at Earth: 1 Initial architecture. *Space Weather*, *13*, 374–385. https://doi.org/10.1002/2015SW001171

Scherrer, P. H., Schou, J., Bush, R. I., Kosovichev, A. G., Bogart, R. S., Hoeksema, J. T., et al. (2012). The Helioseismic and Magnetic Imager (HMI) Investigation for the Solar Dynamics Observatory (SDO). *Solar Physics*, *275*, 207–227. https://doi.org/10.1007/s11207-011-9834-2

Schrijver, C. J. (2009). Driving major solar flares and eruptions: A review. *Advances in Space Research*, *43*, 739–755. https://doi.org/10.1016/j.asr.2008.11.004

Seehafer, N. (1990). Electric current helicity in the solar atmosphere. *Solar Physics*, *125*, 219–232. https://doi.org/10.1007/BF00158402

Shrauner, J. A., & Scherrer, P. H. (1994). East-west inclination of large-scale photospheric magnetic fields. *Solar Physics*, *153*, 131–141.

Song, Q., Zhang, J., Yang, S.-H., & Liu, Y. (2013). Flares and magnetic non-potentiality of NOAA AR 11158. *Research in Astronomy and Astrophysics*, *13*, 226–238. https://doi.org/10.1088/1674-4527/13/2/009

Sterling, A. C., & Hudson, H. S. (1997). Yohkoh SXT observations of X-ray "dimming" associated with a halo coronal mass ejection. *The Astrophysical Journal Letters*, *491*, L55–L58. https://doi.org/10.1086/311043

Thompson, W. T (2006). Coordinate systems for solar image data. *Astronomy & Astrophysics*, *449*, 791–803. https://doi.org/10.1051/0004-6361:20054262

Tziotziou, K., Georgoulis, M. K., & Liu, Y. (2013). Interpreting eruptive behavior in NOAA AR 11158 via the region's magnetic energy and relative-helicity budgets. *The Astrophysical Journal*, *772*, 115. https://doi.org/10.1088/0004-637X/772/2/115

Ulrich, R. K., & Boyden, J. E. (2006). Carrington coordinates and solar maps. *Solar Physics*, *235*, 17–29.

Ulrich, R. K., & Tran, T. (2013). The global solar magnetic field—Identification of traveling, long-lived ripples. *The Astrophysical Journal*, *768*, 189. https://doi.org/10.1088/0004-637X/768/2/189

Ulrich, R. K., & Tran, T. (2016). Generation of a north/south magnetic field component from variations in the photospheric magnetic field. *Solar Physics*, *291*, 1059–1076. https://doi.org/10.1007/s11207-016-0882-5

Wang, Y.-M. (2013). On the strength of the hemispheric rule and the origin of active-region helicity. *The Astrophysical Journal Letters*, *775*, L46. https://doi.org/10.1088/2041-8205/775/2/L46

Zhao, J., Li, H., Pariat, E., Schmieder, B., Guo, Y., & Wiegelmann, T. (2014). Temporal evolution of the magnetic topology of the NOAA active region 11158. *The Astrophysical Journal*, *787*, 88. https://doi.org/10.1088/0004-637X/787/1/88